# SHG (532 nm)-induced spontaneous parametric downconversion noise in 1064 nm-pumped IR upconversion detectors

L. MENG,[1,*,†] A. PADHYE,[2,*,†] C. PEDERSEN,[1] M. EBRAHIM-ZADEH,[2,3] AND P. J. RODRIGO[1,*]

[1]*DTU Fotonik, Technical University of Denmark, Frederiksborgvej 399, DK-4000 Roskilde, Denmark*
[2]*ICFO—Institut de Ciencies Fotoniques, The Barcelona Institute of Science and Technology, 08860 Castelldefels, Barcelona, Spain*
[3]*Institucio Catalana de Recerca i Estudis Avancats (ICREA), Passeig Lluis Companys 23, 08010 Barcelona, Spain*
*Corresponding authors: licme@fotonik.dtu.dk; anuja.padhye@icfo.eu; pejr@fotonik.dtu.dk
†Co-first authors with equal contribution.

**As a novel technique for infrared detection, frequency upconversion has been successfully deployed in many applications. However, investigations into the noise properties of upconversion detectors (UCDs) have also received considerable attention. In this letter, we present a new noise source – second harmonic generation (SHG)-induced spontaneous parametric downconversion (SPDC) – experimentally and theoretically shown to exist in short-wavelength-pumped UCDs. We investigate the noise properties of two UCDs based on single-pass 1064 nm-pumped periodically poled LiNbO$_3$ bulk crystals. One UCD is designed to detect signals in the telecom band and the other in the mid-infrared regime. Our experimental demonstration and theoretical analysis reveal the basic properties of this newly discovered UCD noise source, including its dependence on crystal temperature and pump power. Furthermore, the principle behind the generation of this noise source can also be applied to other UCDs, which utilize nonlinear crystals either in waveguide form or with different bulk materials. This study may also aid in developing methods to suppress the newly identified noise in future UCD designs.**

Recently, frequency upconversion has been demonstrated as a promising infrared (IR) detection method for various applications [1-4]. It is achieved by mixing the IR signal with a pump field inside a $\chi^{(2)}$ material. This sum-frequency generation (upconversion) process spectrally translates the IR signal into a corresponding visible/near-IR output, which enables high-performance detection via silicon (Si) based detectors. Thus, measuring the upconverted (output) signal usually results in a higher signal-to-noise ratio than that of measuring the IR signal directly. The upconversion detector (UCD) offers salient advantages over a direct IR detector (e.g. HgCdTe detector), namely, high efficiency, low background noise, fast response time and room-temperature operation.

In order to achieve high upconversion efficiency, $\eta_{up}$, a high intensity pump field in the $\chi^{(2)}$ material is required. However, this usually leads to unintended nonlinear optical processes, i.e. additional UCD noise sources, such as upconverted spontaneous parametric down-conversion (USPDC) [5,6], upconverted spontaneous Raman scattering [7], and upconverted thermal radiation [8]. It is necessary to emphasize that USPDC exists only in UCDs that utilize a short-wavelength pump ($\lambda_p < \lambda_{IR}$, where $\lambda_p$ and $\lambda_{IR}$ are the wavelengths of the pump and the IR signal, respectively). Especially when quasi-phase-matching (QPM) devices are used, the normally phase-mismatched SPDC process is enhanced due to the random duty-cycle (RDC) error in the periodically structured QPM material [9,10], thereby increasing USPDC noise. For cases in which USPDC is the dominant noise source, the typical dark-count-rate (DCR) of the UCD is larger than $10^5$ /s [6]. Pelc *et al.* suggested the use of long-wavelength pump ($\lambda_p > \lambda_{IR}$) in UCDs to avoid USPDC noise, and they managed to significantly reduce the DCR to a level of $\sim 10^2$ /s [11], which is attributed to the upconverted spontaneous Raman scattering.

Even though the short-wavelength-pumped UCD has a higher DCR, it can still be a better choice for some specific applications given the following advantages it possesses over a long-wavelength-pumped UCD: 1) long-wavelength-pumped UCDs operate only in regions where the IR signal has a shorter wavelength than that of the pump, but for the short-wavelength-pumped UCD, detection can accommodate longer wavelengths and is limited only by the transparency of the nonlinear material

[12]; 2) upconverted output signal wavelength in the long-wavelength-pumped UCD may lie outside sensitive spectral region of the Si detector; 3) long-wavelength-pumped UCDs are usually based on waveguide QPM devices, which have very small étendue [11]. On the other hand, short-wavelength-pumped UCDs have been implemented not only in waveguides but also in bulk nonlinear crystals. UCDs that use bulk nonlinear crystals provide a relatively large étendue, which allows for better collection of the IR signal from a larger solid angle [2].

In this Letter, we focus on the investigation of noise sources for the short-wavelength-pumped UCD. We employed a bulk periodically poled LiNbO$_3$ (PPLN) crystal as the $\chi^{(2)}$ material in our UCD. The PPLN crystal is attractive due to its high nonlinearity, high damage threshold, wide transparent window and large design flexibility. A high-power 1064 nm cw laser is used as the optical pump in our experiment – 1064 nm is a commonly used pump wavelength in PPLN based frequency conversion devices. We measured the parasitic noise of a UCD which is designed for either the telecom or the mid-infrared (MIR) spectral range, respectively. We characterized the properties of the USPDC and the upconverted background thermal radiation. Essentially, the presence of another UCD noise source, second harmonic generation (i.e. 532 nm)-induced SPDC (SHG-SPDC), is identified experimentally for the first time. SHG-SPDC is found to have a significant impact on the overall DCR of the UCD.

In principle, the process of frequency upconversion inside the PPLN crystal involves pump photons of wavelength $\lambda_p$ and IR photons of wavelength $\lambda_{IR}$, which are annihilated to produce upconverted photons of wavelength $\lambda_{up}$. This process follows the principle of energy conservation ($1/\lambda_p + 1/\lambda_{IR} = 1/\lambda_{up}$), and its efficiency is maximized when the k-vectors of the upconverted, IR signal and pump photons fulfill the QPM condition: $k_{up} - k_{IR} - k_p = 2\pi q/\Lambda$, where $\Lambda$ is the poling period of the PPLN crystal and q is an integer indicating the QPM order. However, phase-mismatched parametric processes also exist and can be enhanced due to domain-disorder-induced QPM pedestal caused by the RDC error in the PPLN crystal – one example is USPDC, which has been shown as a prevalent noise source in UCDs [6,11].

Theoretically, several types of parametric processes can occur in strongly pumped QPM materials since photons with different wavelengths can be either upconverted, frequency-doubled or downconverted, although not necessarily phase-matched. Moreover, parametric processes with multiple steps are also possible, which make the noise generation process even more complicated. However, it is unnecessary to analyze all these parametric processes individually since most of them are extremely weak. Primarily, only a single-step process (e.g. upconversion, SPDC) and a two-step process with one step fulfilling the QPM condition (e.g. USPDC) have been considered so far. In this study, we found another two-step process – the unintended SHG of the pump followed by a quasi-phase-matched SPDC (SHG-SPDC) – as another substantial noise source for the UCD. To our best knowledge, this is the first time that the SHG-SPDC process is identified as a noise source for the short-wavelength-pumped UCD. It can potentially contribute to the overall background noise of short-wavelength-pumped UCDs for a number of applications [6,12–17], especially if the bandpass filters used to collect the upconverted output are not spectrally narrow enough. Most importantly, the principle of this specific noise source also applies to other QPM devices (e.g. PPLN-waveguide, PPKTP and orientation-patterned GaP). We believe that this noise source should be carefully considered in future UCD applications.

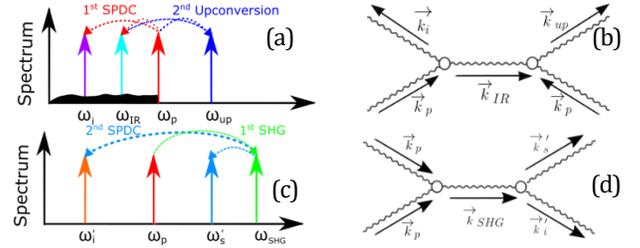

Fig. 1. Respective spectral and Feynman diagrams for the USPDC (a, b) and the SHG-SPDC (c, d) noise generation process. $\omega_j = 2\pi c/\lambda_j$.

For comparison, the spectral and Feynman diagrams for the USDPC and the SHG-SPDC are shown in Figs. 1(a-d). Note that, both respective 1st steps (SPDC in USPDC and SHG in SHG-SPDC) are phase-mismatched. Therefore their intensities are several orders of magnitude lower than that of the pump field. The SPDC in USPDC is broadband due to the QPM pedestal caused by the RDC error [9,10]. In contrast, the SHG is narrowband due to the typically narrow linewidth of the pump. The 2nd steps (the upconversion in USPDC and the SPDC in SHG-SPDC) are quasi-phase-matched. It is necessary to emphasize that the 1st-order QPM condition (q = 1) for the upconversion is deliberately fulfilled by the chosen poling period at the intentionally maintained crystal temperature range. On the other hand, the QPM condition for the SPDC in SHG-SPDC, $k_{SHG} - k_s' - k_i' = 2\pi q/\Lambda$, is accidentally fulfilled, but only for certain $\Lambda$, q and operating temperature values.

Figure 2 shows a sketch of the single-pass 1064 nm-pumped upconversion system. The experimental setup can be separated into three parts (marked by dashed squares). Note that, the IR collection part (yellow square) is not included in our experiment since here we focus only on the noise investigation, but it can be installed in order to use this system as an IR detector. In the upconversion part (red square), a single-frequency, cw Yb-fiber laser (YLR-30-1064-LP-SF, IPG Photonics) with a maximum output power of 30 W is used as the 1064 nm pump for upconversion. After the collimation, the output of the fiber laser is focused by a lens L1 (f1 = 200 mm) into a PPLN crystal (1/e$^2$ beam waist radius $w_0$ = 48 μm). The PPLN crystal is mounted in a PPLN oven (PV40, Covesion) with the ability of temperature tuning. After passing through the PPLN crystal, the 1064 nm pump is reflected by a dichroic mirror (DM2) into a power meter. The noise photons generated inside the PPLN crystal pass through DM2 and enter the noise detection part (blue square). After passing through a periscope (M1, M2), the noise photons change to the p-polarization state with respect to the working surface of a prism (N-SF11) used to separate the noise from the 1064 nm pump. An iris is placed in the image plane of L2 (f2 = 50 mm) in order to filter out the residual pump and its 532 nm SHG beam. For further reduction of the ambient light, a bandpass filter is placed in front of a sensitive EM-CCD camera (Luca S 658M, Andor), which is used to measure the amount of noise photons. The noise photon rate (i.e. DCR) is calculated as $(R - R_0)/\tau$, where R and $R_0$ are the sums of the camera pixel readout in the region of interest at $P_p > 0$ and $P_p = 0$, respectively; $\tau$ is the integration time of the camera. In order to use the full dynamic range of the camera, $\tau$ is selected from 3 sec to 1 min according to the intensity of the noise. Note that, two different multiple poling period 5% MgO-doped PPLN crystals (one suitable for telecom and the other for MIR detection) are used in our experiment, and appropriate bandpass filters are used accordingly.

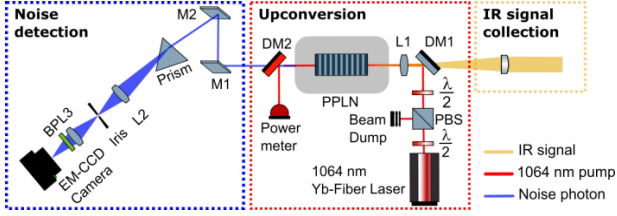

Fig. 2. Experimental setup for the noise measurement. λ/2, half-wave plates; PBS, polarization beam splitter; DM1-2, dichroic mirrors; L1-3, lenses with focal lengths f1 = 200 mm, f2 = 50 mm, and f3 = 50 mm, respectively; M1-2, mirrors; BP, bandpass filters.

First, the noise from a 40-mm long PPLN crystal with two channels of Λ = 12 µm and Λ = 12.69 µm is measured separately. The corresponding wavelengths of the noise photons ($\lambda_s$' and $\lambda_{up}$) are calculated based on the QPM conditions for the SPDC of 532 nm (dashed lines, $\lambda_{SHG} \rightarrow \lambda_s' + \lambda_i'$) and the upconversion (solid lines, $\lambda_{IR} + \lambda_p \rightarrow \lambda_{up}$) as seen in Fig. 3(a). In the temperature range of 20 °C to 100 °C, $\lambda_s$' and $\lambda_{up}$ are in the range between 600 nm and 650 nm. A bandpass filter (at $\lambda_c$ = 625 nm with bandwidth $\Delta\lambda_{FWHM}$ = 50 nm) is used to remove the ambient light outside the spectrum of interest. Figures 3(b-c) show the images taken by the EM-CCD camera while the temperature of the crystal is kept at 50 °C. The images suffer from astigmatism due to dispersion of the prism.

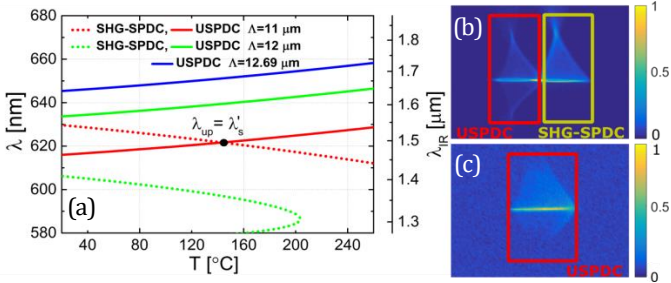

Fig. 3. (a) Wavelength ($\lambda_s$' and $\lambda_{up}$) versus temperature given by the QPM conditions for the SPDC in SHG-SPDC and the upconversion in PPLN. Images used for the noise measurement at crystal temperature T = 50 °C using poling periods (b) Λ = 12.0 µm and (c) Λ = 12.69 µm.

For the poling period of Λ = 12 µm, the wavelength of the SHG-SPDC and the USPDC noise can be calculated as $\lambda_s$' = 605 nm and $\lambda_{up}$ = 634.9 nm, respectively. Therefore, these two kinds of noise photons can be separated spatially by the prism, as shown by two diamond-shaped patterns in Fig. 3(b). The USPDC and the SHG-SPDC parts are marked by red and yellow boxes, respectively. The case of Λ = 12.69 µm in Fig. 3(c), only has one diamond-shaped pattern, which is caused by the USPDC process ($\lambda_{up}$ = 646.7 nm). The other pattern disappears due to the phase-mismatch of the SHG-induced SPDC process in this case. From the perspective of the UCD, these two poling periods can have similar background noise levels since the noise photons due to SHG-SPDC for the poling period of Λ = 12 µm can be removed easily using a narrow bandpass filter. However, in some particular cases, the noise due to SHG-SPDC cannot be removed since $\lambda_s$' and $\lambda_{up}$ can have the same value, as shown by the black dot in Fig. 3(a), which implies that if Λ = 11 µm is used with a crystal temperature of 147 °C, the UCD will suffer from a higher total background noise.

Next, the respective DCRs due to the USPDC and the SHG-SPDC are measured as a function of varying pump power ($P_p$) for the poling period of Λ = 12 µm while the operating temperature of the PPLN crystal is kept fixed at 50 °C [see Fig. 4(a)]. At $P_p$ = 22 W, we theoretically estimate $\eta_{up}$ to be close to unity. The SPDC intensity is known to be proportional to $P_p$ [10] while $\eta_{up}$ is linear in $P_p$ for $\eta_{up}$ well below 0.5 but saturates as it approaches unity [11]. Thus, the USPDC rate is proportional to $P_p^2$ at first but behaves linearly as $P_p$ approaches a value where $\eta_{up}$ is maximum, as seen from the experimental data fitted with a 2$^{nd}$-degree polynomial [Fig. 4(a)]. The SHG-SPDC data is fitted with 2$^{nd}$- and 3$^{rd}$-degree polynomials [pink dashed and solid curves, respectively], indicating a better fit, i.e. a lower reduced chi-squared, for the latter. In a solitary SHG process, the SHG rate is known to scale with $P_p^2$. However, the fit of SHG-SPDC data to a 3$^{rd}$-degree polynomial in $P_p$ suggests that the SHG and SPDC processes are coupled in a more complex manner (e.g. SHG rate increases along the crystal length). A more rigorous model is required to explain the dependence of SHG-SPDC on $P_p$.

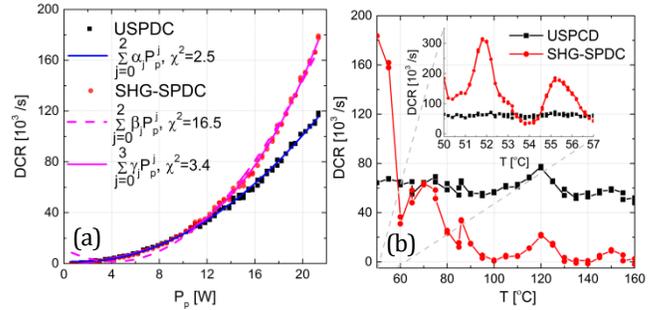

Fig. 4. Measured DCRs caused by the USPDC and the SHG-SPDC as a function of (a) 1064 nm pump power and (b) operating temperature of the PPLN crystal (See Visualization 1 for the series of images taken at different temperatures). $\alpha_j$, $\beta_j$ and $\gamma_j$ are the fitting parameters. The uncertainty of each measured DCR is less than $4 \times 10^3$ /s.

Using the same poling period (Λ = 12 µm), the DCR due to the USPDC and the SHG-SPDC are measured again but with different crystal temperatures at 1064 nm pump power of 22 W. Figure 4(b) shows the measurement results, which imply that the change in temperature (from 50 °C to 57 °C) has a larger impact on the SHG-SPDC process than on the USPDC process. We think this is due to the significant fluctuation of the SHG intensity when the UCD operates at different crystal temperatures. In general, these two noise sources have similar DCR levels, but at a certain temperature (52 °C), the DCR due to the SHG-SPDC can be 5 times that due to the USPDC. Therefore, SHG-SPDC noise should be avoided in practical applications. Otherwise, it may become the dominant UCD noise source. When the temperature is higher than 100 °C, $\lambda_s$' becomes lower than 600 nm, and the bandpass filter starts blocking it. Thus, the measured DCR in Fig. 4(b) due to SHG-SPDC process is strongly reduced for T > 100 °C. The PPLN crystal with poling periods of Λ = 12 µm and Λ = 12.69 µm are suited for

IR detection in the telecom band (1.5 μm < $\lambda_{IR}$ < 1.7 μm), where the upconverted thermal radiation can be neglected. In contrast, when the UCD operates in the MIR region, the upconverted thermal radiation (mainly from the PPLN crystal) becomes the dominant noise source [8]. In the following experiment, a 20-mm long PPLN crystal with two channels of Λ = 22 μm and Λ = 23 μm (for MIR detection) is used for the noise analysis. The wavelength of the upconverted signal $\lambda_{up}$ is between 750 nm and 850 nm. Accordingly, a new bandpass filter with $\lambda_c$ = 800 nm and $\Delta\lambda_{FWHM}$ = 100 nm is used here to filter out the ambient light. Figures 5(b-c) show the images taken by the camera when the temperature of the PPLN crystal is kept fixed at 50 °C. Similar to Fig. 3(b), Figs. 5(b-c) have two separate patterns, which indicate that both channels fulfill the QPM condition for the SHG-SPDC. It is also predicted by the QPM curve in our theoretical analysis [see Fig. 5(a)]. But it is necessary to emphasize that it is the 3$^{rd}$-order QPM that is used for the SHG-SPDC here, i.e. $k_{SHG} - k_s' - k_i' = 3 \cdot 2\pi/\Lambda$. Therefore, the intensity of the SHG-SPDC, in this case, is smaller than that of the accompanying USPDC. The red boxes in Figs. 3(b-c), i.e. UCD for telecom, are labelled only as "USPDC" since the noise originates primarily from the USPDC process. In contrast, the red boxes in Figs. 5(b-c) are labelled as "Thermal & USPDC" since the noise contribution from the upconverted thermal radiation cannot be neglected here. In fact, it even becomes the dominant noise source for the case of Λ = 23 μm and $\lambda_{IR}$ is larger than 4 μm [8].

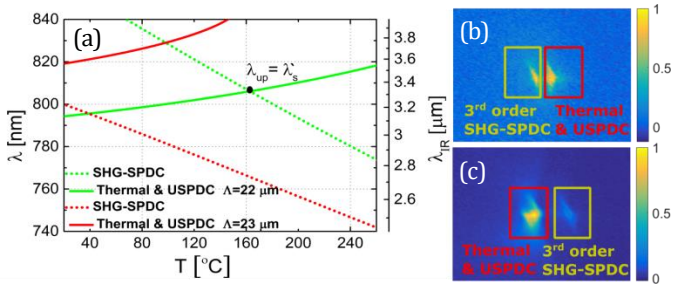

Fig. 5. (a) Wavelength ($\lambda_s'$ and $\lambda_{up}$) versus temperature given by the QPM conditions for the SPDC in SHG-SPDC and the upconversion in PPLN. Images used for the noise measurement at crystal temperature T = 50 °C using poling periods (b) Λ = 22.0 μm and (c) Λ = 23.0 μm.

The black dot in Fig. 5(a) [similar to the black dot in Fig. 3(a)] is the intersection point between the green solid and the green dashed lines (Λ = 22 μm). It implies that the two noise sources have the same wavelength ($\lambda_s' = \lambda_{up}$) at a PPLN crystal temperature of 165 °C. In order to show this wavelength degeneracy, a series of images is taken at different crystal temperature settings. An overlap of the two noise patterns is experimentally observed when the temperature is around 160 °C (See Visualization 2). For Λ = 23 μm, the spatial separation between the two patterns increases with crystal temperature (See Visualization 3), which is consistent with the theoretical calculation [red solid and dashed lines in Fig. 5(a)]: the spectral separation increases with operating temperature.

Pumped by a 1064 nm laser, bulk or waveguide PPLN crystals with poling periods around 12 μm and 20 μm are used in several applications requiring telecom and MIR UCDs, respectively [6,12–17]. However, similar to what we demonstrate here, these applications are in spectral regions where the SHG-SPDC process can potentially introduce additional UCD noise. Moreover, the spectral separation between the signal and the noise may be too small for the noise to be removed by narrow bandpass filters.

In summary, we built a single-pass, short-wavelength-pumped UCD, and investigated its noise properties when operating in either the telecom or the MIR range. Most notably, we discovered a new noise source, SHG-SPDC, which can potentially increase the noise floor (i.e. DCR) of the UCD. Similar to the USPDC process, the SHG-SPDC noise generation is also a two-step process: phase-mismatched SHG is accompanied by a quasi-phase-matched SPDC. Experimentally, its intensity is found to be approximately proportional to the cube of the pump power. Our experimental results show that this particular noise source usually has a different wavelength than the upconverted signal, so it can be removed by using spectral filters. However, for some particular cases, SHG-SPDC noise may have the same wavelength as the upconverted signal, which makes spectral filtering impossible. The results presented in this paper can also be extended to other forms of UCD which may use waveguide configuration, different nonlinear materials, or different pump wavelengths. This newly discovered noise source should be carefully considered in the design and operation of future UCDs, particularly those intended for weak (e.g. single-photon level) IR signal detection.

**Funding.** Mid-TECH – H2020-MSCA-ITN-2014 (642661). Severo Ochoa (SEV-2015-0522-16-1); Fundación Cellex.

**Acknowledgment**. We thank P. Tidemand-Lichtenberg, L. Høgstedt, K. Devi, and C. K. Suddapalli for fruitful discussions.